\documentclass[aps,prb,twocolumn,superscriptaddress,showpacs]{revtex4}

\usepackage{textcomp}
\usepackage{amsmath}
\usepackage{graphicx}
\usepackage{color}

\newcommand{\jmmapprox}{\sim \! \!}

\begin{document}

% ******************************
% Title of paper
% ******************************
\title{Towards a Predictive First-Principles Description of Solid Molecular Hydrogen with Density-Functional Theory}

% ******************************
% Authors / affiliations
% ******************************
\author{Miguel A. Morales}
\email[]{moralessilva2@llnl.gov}
\affiliation{Lawrence Livermore National Laboratory, Livermore, California 94550, USA}

\author{Jeffrey M. McMahon}
\affiliation{Department of Physics, University of Illinois at Urbana-Champaign, Urbana, Illinois 61801, USA}

\author{Carlo Pierleoni}
\affiliation{Department of Physical and Chemical Sciences, University of L'Aquila and CNISM UdR L'Aquila, Via Vetoio, I-67010 L'Aquila, Italy}

\author{David M. Ceperley}
\affiliation{Department of Physics, University of Illinois at Urbana-Champaign, Urbana, Illinois 61801, USA} 

% ******************************
% Date
% ******************************
\date{\today}

% ******************************
% Abstract
% ******************************
\begin{abstract}
We examine the influence of the main approximations employed in density-functional theory descriptions of the solid phase of molecular hydrogen near dissociation. We consider the importance of nuclear quantum effects on equilibrium properties and find that they strongly influence intramolecular properties, such as bond fluctuations and stability. We demonstrate that the combination of both thermal and quantum effects make a drastic change to the predicted optical properties of the molecular solid, suggesting a limited value to dynamical, e.g., finite-temperature, predictions based on classical ions and static crystals.
We also consider the influence of the chosen exchange--correlation density functional on the predicted properties of hydrogen, in particular, the pressure dependence of the band gap and the zero-point energy. Finally, we use our simulations to make an assessment of the accuracy of typically employed approximations to the calculation of the Gibbs free energy of the solid, namely the quasi-harmonic approximation for solids. We find that, while the approximation is capable of producing free energies with an accuracy of $\approx 10$ meV, this is not enough to make reliable predictions of the phase diagram of hydrogen from first-principles due to the small free energy differences seen between several potential structures for the solid; direct free energy calculations for quantum protons are required in order to make definite predictions.  
\end{abstract}


%\keywords{}

\maketitle

%%%%%%%%%%%%%%%%%%%%%%%%%%%%%%%%%%%%%%%%%%%%%%%%%%
\section{Introduction}
%%%%%%%%%%%%%%%%%%%%%%%%%%%%%%%%%%%%%%%%%%%%%%%%%%

Hydrogen at high pressure, being the most abundant element in the Universe, plays a prominent role in many scientific fields, notably in modeling of the giant planets. Its simplicity, i.e., being comprised of a single proton and electron per atom,  makes it a unique system from a theoretical and computational standpoint. Thus, considerable attention has been devoted to understanding hydrogen both experimentally and theoretically, as has been recently reviewed \cite{McMahon12}.

Numerous exciting experimental breakthroughs have recently been made in the low-temperature solid phase of molecular hydrogen, including a controversial observation of metallization \cite{H_metallization_Eremets-NatMat-2011}  as well as the discovery of a new phase, Phase IV \cite{Howie12}. Despite these advances, many open questions remain, such as whether metallization has actually been achieved \cite{Nellis12} and the structure of this new phase \cite{Howie12}. Computational predictions have accompanied these results \cite{Pickard12,Liu12}. However, the agreement  of these predictions with experiment is not perfect, and  have not resolved these questions.

To help resolve these discrepancies, a careful examine of the approximations made in such simulations is neededd. For example, in almost all studies of hydrogen reported to date, especially in the solid phase, many properties have been calculated for perfect lattices \cite{Ceperley87,Natoli93,Natoli95},  with a finite-temperature description based on classical protons \cite{Bonev04,Morales10}, or within a quasi-harmonic calculation \cite{Pickard07,McMahon11,Pickard12,Liu12}. While such procedures would be justified if both nuclear quantum effects (NQEs) and thermal fluctuations were small, such is not the case for hydrogen at the high pressures under consideration.  Notice that even at temperatures below $T = 100$ K, the kinetic energy  of the protons in the crystal is on the order $\jmmapprox 1000$ K, because of proton zero point energy (ZPE).
While some progress has been made in the direct treatment of NQEs in hydrogen from first principles recently \cite{Geneste12,Morales13,McMahon13}, there is still considerable work to be done. 

Another major approximation arises in density-functional theory (DFT) studies of hydrogen, namely the choice of the approximate exchange--correlation density functional (DF). Local or semi-local DFs, such as that of Perdew--Burke--Ernzerhof (PBE) \cite{Perdew96}, have been the standard choice in DFT simulations over the last decade. However, it has recently been demonstrated, at least at higher temperature in the liquid phase \cite{Morales13}, that the use of nonlocal functionals, such as that by Heyd--Scuseria--Ernzerhof (HSE) \cite{Heyd03}, which contains a fraction of exact exchange or one with an improved description of dispersion interactions, such as vdW-DF2, significantly improve the description of the dissociation process. Since the neglect of NQEs and the deficiencies of the PBE DF partially compensate each other in many situations\cite{Morales13}, most calculations to date have made the assumption that this cancellation is accurate. As we show below, accurate and predictive simulations of hydrogen, in particular in the region where molecular dissociation and metallization occur, require a more rigorous treatment of electronic exchange and correlation effects (i.e., beyond those provided by typical semi-local functionals like PBE). %In addition, most properties of the solid are strongly affected by NQEs, leading to very poor predictions when these effects are not properly taken into account. 
%This is particularly important in the prediction of the correct structure and in the prediction of optical and vibrational properties as a function of pressure. %  \jmm{Discuss the structure being of upmost importance, as well as the band gap which can be very important in this region and near metallization that can drive phase transitions.}

%\jmm{Emphasize that we are not addressing these recent predictions, etc. in this article, but rather trying to examine the theoretical methods themselves that are used to make the predictions} 
The main purpose of this article is to examine the effect of these approximations in first-principles simulations of crystalline molecular hydrogen, in order to improve the accuracy of the method. We focus our study on the influence of NQEs and the choice of DF on the orientational order in the crystal and on bandgaps. We also make a critical assessment of some common approximations, such as the quasi-harmonic approximation applied to the prediction of the relative energies of competing structures. We begin in Section \ref{sec:comp_details} by providing the computational details for the results that follow in Section \ref{sec:results}. Section \ref{sec:discussion} provides a brief discussion of the results, and Section \ref{sec:concl} concludes.

%%%%%%%%%%%%%%%%%%%%%%%%%%%%%%%%%%%%%%%%%%%%%%%%%%
\section{Computational Details}
\label{sec:comp_details}
%%%%%%%%%%%%%%%%%%%%%%%%%%%%%%%%%%%%%%%%%%%%%%%%%%

We performed first-principles simulations of hydrogen using DFT.  Three DFs were considered: the semi-local PBE DF \cite{Perdew96}, the HSE DF \cite{Heyd03}, containing a fraction of exact-exchange, and the vdW-DF2 DF \cite{Dion04,Thonhauser07,Roman-Perez09,Lee10}, capable of treating van der Waals (vdW) interactions. Simulations using HSE were performed with a modified version of VASP \cite{VASP}, while those with PBE and vdW-DF2 were performed with a modified version of Quantum ESPRESSO (QE) \cite{QE}. A Troullier--Martins norm-conserving pseudopotential \cite{Troullier91} with a core radius of $r_c = 0.5$ a.\ u.\ was used to replace the bare Coulomb-potential of hydrogen in the QE simulations and a PAW \cite{Kresse99} was used in VASP. Planewave cutoffs of 1224 eV and 350 eV were used in these simulations, respectively.

Path-integral molecular dynamics (PIMD) simulations were employed via the accelerated method of Ceriotti, \emph{et al.} \cite{Ceriotti}, based on a generalized Langevin dynamics (GLE) and the Born--Oppenheimer (BO) approximation, which we indicate as PI+GLE. The use of the PI+GLE method was carefully tested under the relevant pressure and temperature conditions, in order to guarantee proper convergence \cite{McMahon12b}.
A time-step of 8 (a.\ u.)$^{-1}$ was used in all simulations, and the PIs were discretized with a Trotter time-step no larger than $0.0003125$ K$^{-1}$. After an equilibration period of $\jmmapprox 0.25$ ps, statistics were gathered for simulation times of $\jmmapprox 1.25$--$1.75$ ps, corresponding to $\jmmapprox 6500$--$9000$ time steps. 
A $2^3$ \textbf{k}-point grid was used in simulations with the PBE and vdW-DF2, while a $1 \times 2 \times 2$ grid was used in simulations with HSE \footnote{The unit cells are elongated along the x axis, so a 1x2x2 \textbf{k}-point grid provided a reasonable compromise between speed and accuracy in simulations with HSE.}. Finite-temperature effects on the electrons were taken into account using Fermi--Dirac smearing \cite{Kresse99}. 

We performed both classical nuclei and quantum simulations for all DFs of three of the 
primary candidate structures for the high-pressure molecular phases (in the region of $\jmmapprox 300$ GPa) \cite{Pickard07}: $C2c$, 
$Cmca$-$12$, and $Pbcn$. All calculations were performed in the $NVT$ ensemble, where $N$ is the number of particles and $V$ is the volume. Simulation cells contained $144$ atoms at temperatures between $200$ K and $500$ K with pressures from $200$--$550$ GPa. 
Note that below, we use the term BOMD to refer to simulations that treat the protons as classical particles, while PIMD refers to those including a full path integral treatment of the protons (except for quantum statistics of the protons).  
Bandgaps, at finite temperature, were calculated by performing excited-state calculations on $15$ statistically-independent proton configurations sampled from the trajectories. Since semi-local DFs  are well known to underestimate the bandgap \cite{stadele00}, unless otherwise stated, the HSE functional was used to calculate bandgaps. In other words, the trajectories and optical properties were not necessarily calculated using the same DF, and for simplicity, in our descriptions, the label refers to the DF used to generate the trajectories.

%%%%%%%%%%%%%%%%%%%%%%%%%%%%%%%%%%%%%%%%%%%%%%%%%%
\section{Results}
\label{sec:results}
%%%%%%%%%%%%%%%%%%%%%%%%%%%%%%%%%%%%%%%%%%%%%%%%%%

%--------------------------------------------------
\subsection{Orientational-order}
%--------------------------------------------------

We begin by considering perhaps the property of prime importance, the crystal structure of the solid at finite temperature. In particular we look at its orientational order. Since most studies to date have employed PBE in DFT studies \cite{McMahon12}, we use this choice initially to examine NQEs, but later we will assess the influence of other DFs. 

\begin{figure}[t]
     \includegraphics[scale=0.3]{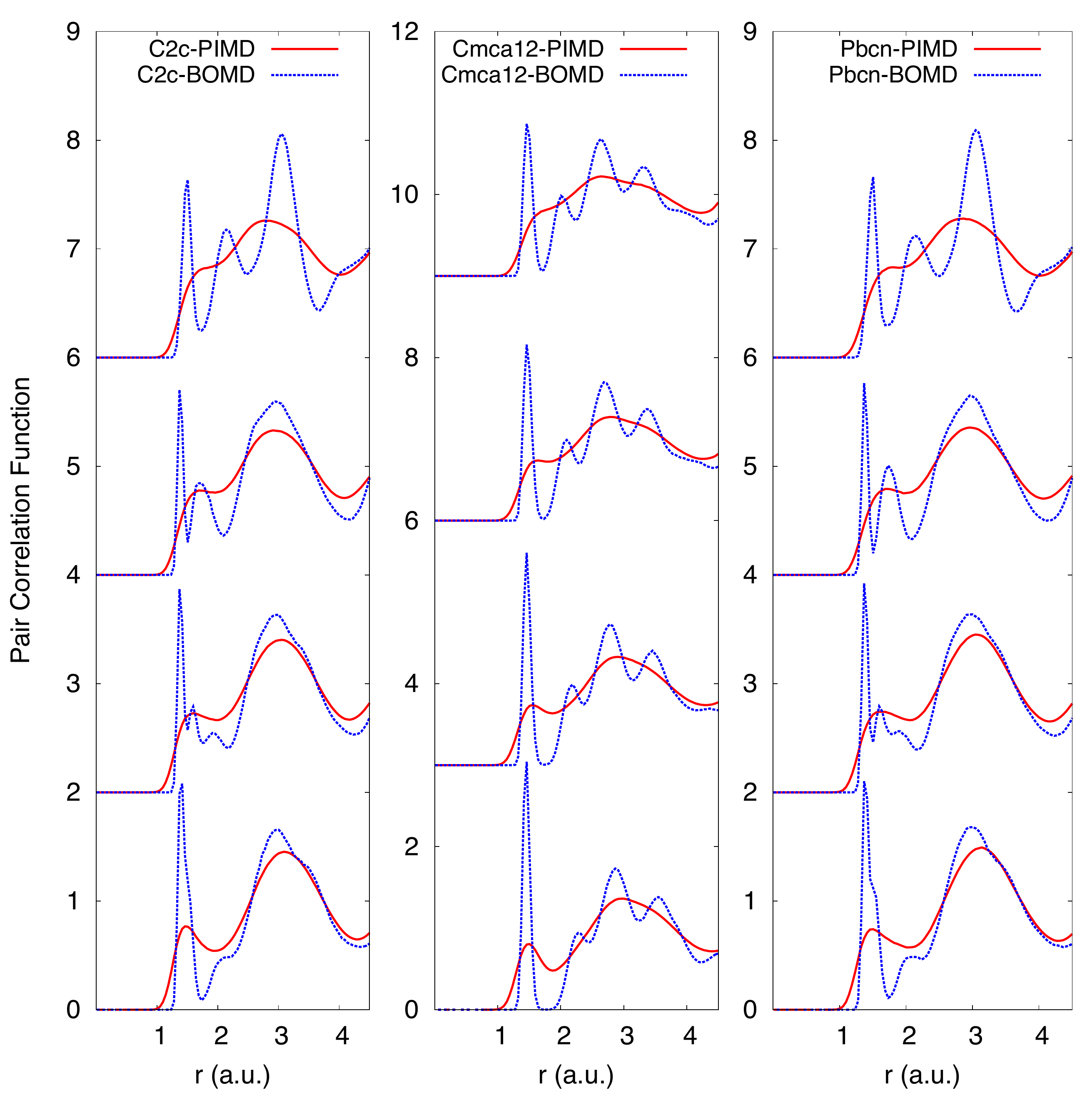}
     \caption{(Color online) The proton--proton PCFs for the $C2c$ (left), $Cmca$-$12$ (center), and $Pbcn$ (right) structures of hydrogen for BOMD (dashed blue) and PIMD (solid red) simulations, using the PBE DF. From top to bottom the PCFs correspond to pressures of $p \jmmapprox$ $350$, $300$, $250$, and $200$ GPa. All simulations were performed at $T = 200$ K.}
     \label{fig:gr_pimdvscl_PBE_allS}
\end{figure}

Figure \ref{fig:gr_pimdvscl_PBE_allS} shows comparisons of the pair correlation functions (PCFs) of hydrogen computed with BOMD and PIMD in
the $C2c$, $Cmca$-$12$, and $Pbcn$ phases. 
A marked disagreement between classical and quantum results is
clearly seen; while classical simulations produce PCFs with considerable structure, including the existence  
of molecules with different bond lengths (as previously reported \cite{Pickard12,Liu12,Labet12_1,Labet12_2,Labet12_3,Labet12_4}), the quantum simulations are seen to have much less structure.

Another interesting feature of Fig.\ \ref{fig:gr_pimdvscl_PBE_allS} is the overlap between neighboring molecules when NQEs are included. In the case of classical simulations, there is a clear separation between close molecules
with almost no overlap with the closest shell. With the inclusion of NQEs though, not only is the height of the molecular
peak dramatically reduced, but the first minimum disappears above pressures of $\jmmapprox 300$ GPa. While this might suggest the dissociation of molecules (i.e., since there is no clear separation between the intramolecular distance
and the nearest-neighbor separation), a further examination of the trajectories suggests stable molecules at all pressures considered.

\begin{figure}[t]
     \includegraphics[scale=0.65]{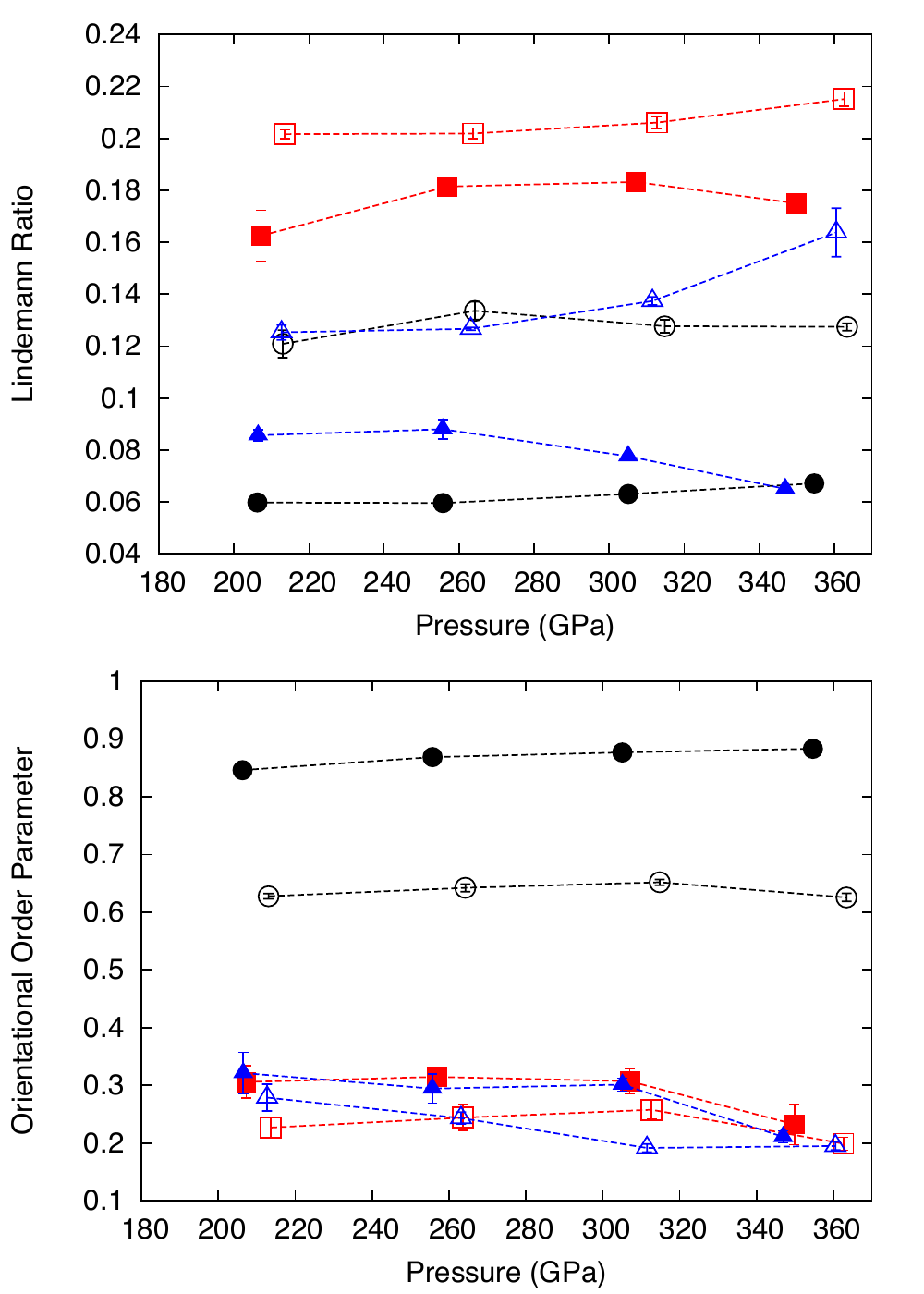}
     \caption{(Color online) Lindemann ratio (top) and orientational order parameter (bottom) of various structures of solid molecular hydrogen from BOMD (solid symbols) and PIMD (open symbols) simulations, using the PBE DF.  All the simulations were performed at a temperature of $T = 200$ K. Squares correspond to $C2c$, circles to $Cmca$-$12$, and triangles to $Pbcn$. } 
     \label{fig:lind_PBE}
\end{figure}

Figure \ref{fig:lind_PBE} shows the Lindemann ratio of the molecular center of mass and the orientational order parameter for both 
BOMD and PIMD simulations. 
There are several ways to measure orientational order \cite{Runge92}; 
we measure the deviation  of  a molecule $\bf{\hat{\Omega}}_i$ during the simulation with respect to the perfect static lattice orientation $\bf{\hat{e}}$:
\begin{equation}
    \hat{O} = \left [ \frac{1}{N} \sum_{i=1}^{N} P_2( \bf{\hat{\Omega}_i} \cdot \bf{\hat{e}}_i) \right ]^2
\end{equation} 
where $P_2$ is the Legendre polynomial. One obtains $\langle \hat{O} \rangle = 1$ in the static lattice, while a solid devoid of orientational order (e.\ g.\ Phase-I of hydrogen \cite{McMahon12}) gives $\langle \hat{O} \rangle = 0$. While the Lindemann ratio depends on structure, as expected, it is fairly insensitive to pressure in the studied range: the molecules are stable and the crystals do not melt. 
Note that $\langle \hat{O} \rangle$ is also insensitive to pressure in this range. 
NQEs are seen to have a moderate effect in the Lindemann ratio of the molecular centers, which suggests that the marked differences observed in the PCFs come largely from strong NQEs on bond fluctuations.

Notice that there is a significant difference in the amount of orientational order between the $Cmca$-$12$ structure and the other two considered when NQEs are included. Molecules in the $Cmca$-$12$ structure largely retain the crystalline orientation even at the temperature investigated. In this case, NQEs are quite consistent, reducing the order parameter by roughly $20\%$. In the other two structures though, the amount of order of the perfect crystal is much more reduced by temperature already, while NQEs play less of a role (a further reduction of only $\jmmapprox 5$--$10\%$). 
This result suggests that NQEs influence structures in different ways and with different magnitudes.

\begin{figure}[h]
     \includegraphics[scale=0.3]{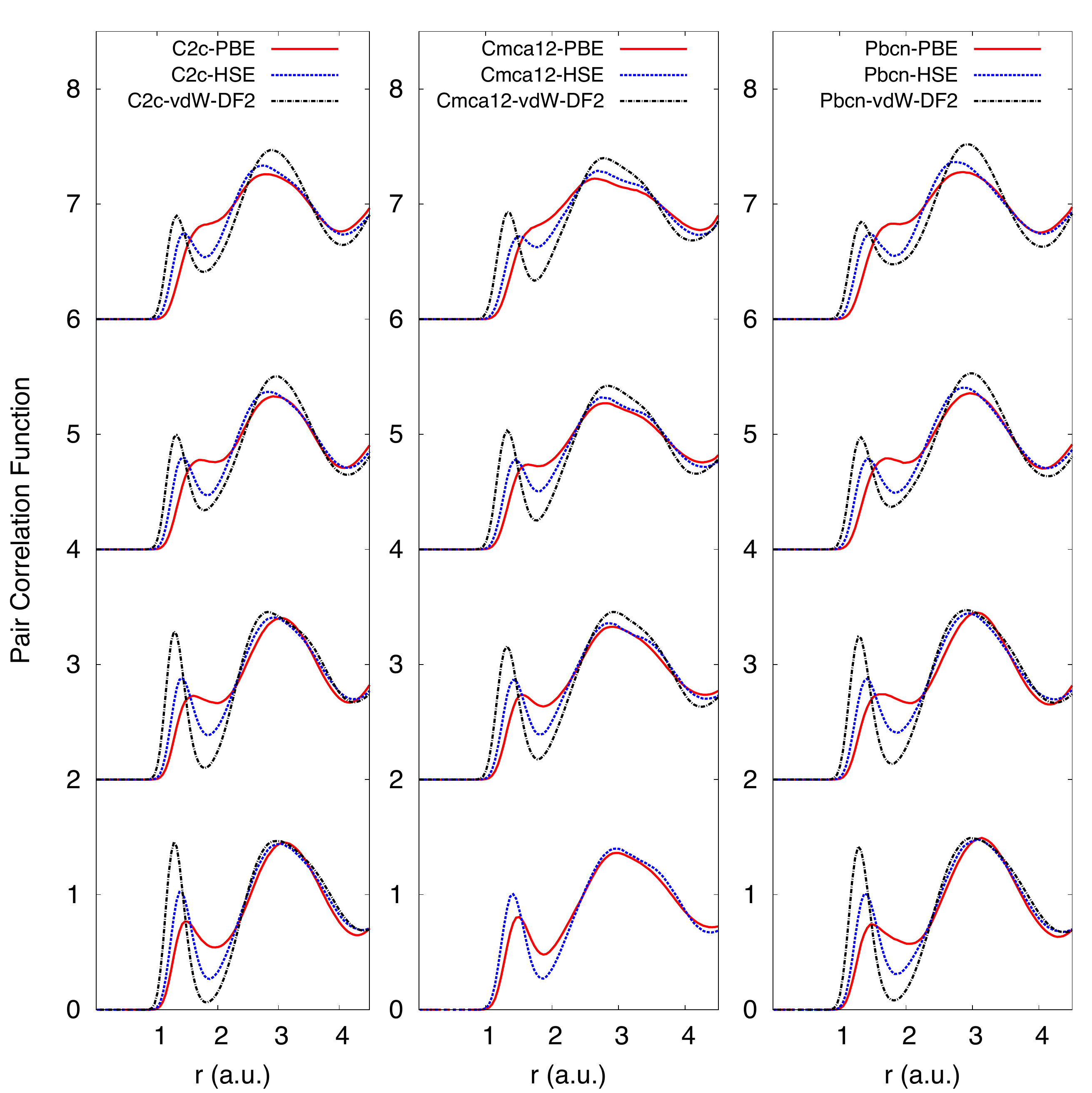}
     \caption{(Color online) PCFs for the $C2c$ (left), $Cmca$-$12$ (center), and $Pbcn$ (right) structures of hydrogen from PIMD simulations using PBE (solid red), HSE (dotted blue), and vdW-DF2 (dashed-dotted black). From top to bottom, the PCFs correspond to pressures of approximately $p \jmmapprox 350$, $300$, $250$, and $200$ GPa. All simulations were performed at $T = 200$ K.  } 
     \label{fig:gr_pimdvscl_allDF_allS}
\end{figure}

\begin{figure}[t]
     \includegraphics[scale=0.65]{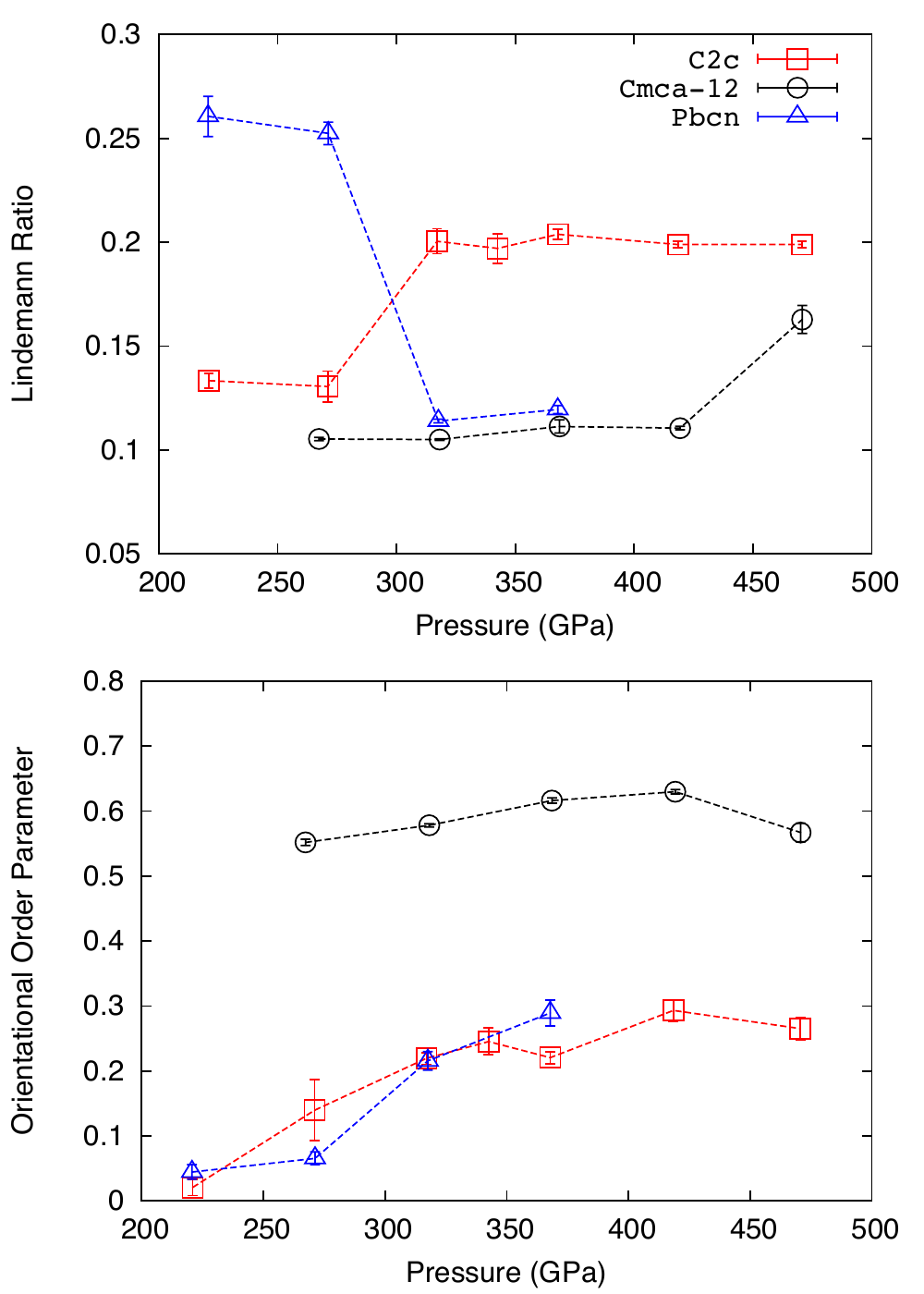}
     \caption{(Color online)  Lindemann ratio (top) and orientational order parameter (bottom) for solid molecular hydrogen from PIMD simulations using vdW-DF2. Note that all the simulations were performed at $T = 200$ K. Squares correspond to $C2c$, circles to $Cmca$-$12$, and triangles to $Pbcn$.
     } 
     \label{fig:lind_vdW}
\end{figure}

Having established the importance of NQEs, it is also important to consider the influence of DF. 
Figure \ref{fig:gr_pimdvscl_allDF_allS} shows a comparison of the PCFs between PBE, HSE, and vdW-DF2, all obtained from PIMD simulations.
Several prominent features are clear. On the one hand, vdW-DF2 simulations
result in structures with more pronounced molecular features, while those with PBE lead to the
structures with less molecular features. 
%This is consistent with the calculated band gaps.
Although, at high pressures, the molecular peak
becomes very weak, molecules do not show any tendency to dissociate.
Similar to the results shown in Fig.\ \ref{fig:lind_PBE}, this can be seen in the Lindemann ratio of 
the molecules and the orientational order parameter, shown in Fig.\ \ref{fig:lind_vdW} for the simulations with vdW-DF2.
A somewhat large, but stable, Lindemann ratio can be seen for $Pbcn$ at the two lowest
pressures, with similar behavior for $C2c$ at the higher pressures. This suggests a 
distortion in these structures at finite temperature, however, the molecules stay intact.
% (\CP{do you mean that the crystal is stable?}).  YES

Note also that the orientational order
parameter goes to zero at pressures below $\jmmapprox 200$ GPa in both the $Pbcn$ and $C2c$ structures,
suggesting a possible transition to Phase-I somewhere between 200 and 250 GPa. 
%\jmm{Is this consistent with experiment, and what is the question mark for here?} MAM: this is too high, the transition is ~ 175 GPa. But bosonic exchange should lower the number closer to experiment
A precise prediction of the location of the I-III phase boundary requires the treatment of bosonic exchange, which we are not considering here.

%--------------------------------------------------
\subsection{Bandgaps}
%--------------------------------------------------

\begin{figure}[ht]
     \includegraphics[scale=0.80]{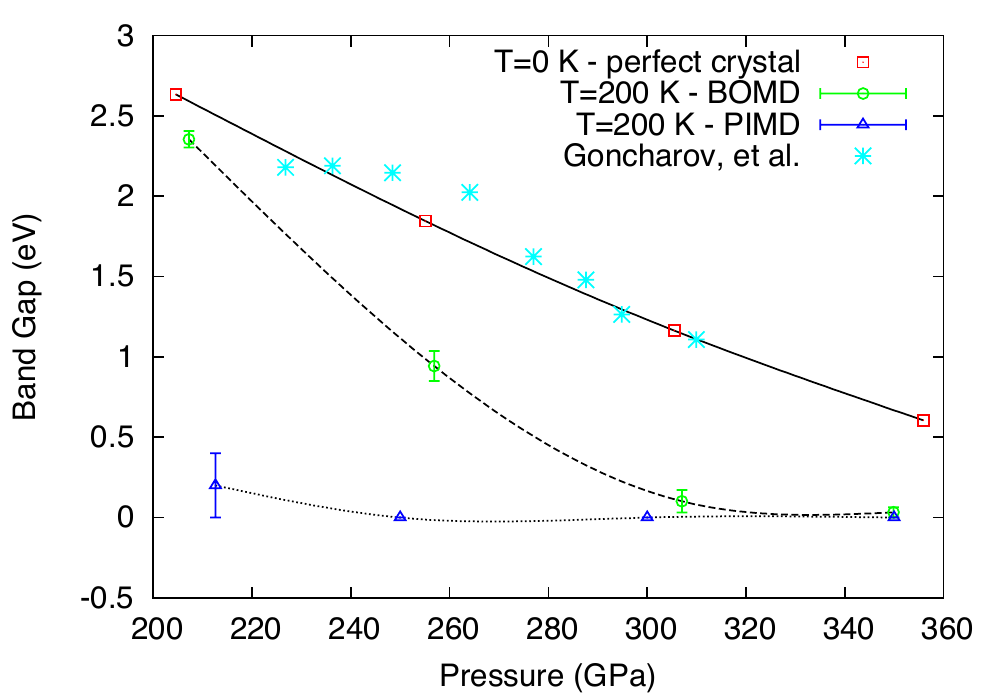}
     \caption{(Color online) Bandgap of $C2c$ using PBE. Red squares are the bandgaps for the static crystal, green circles are band gaps for classical nuclei at 200 K, and blue triangles are band gaps for quantum nuclei at $200$ K. Cyan crosses are recent experimental measurements from Goncharov, \emph{et al.}, \cite{Goncharov12}. Only one set of experimental results is shown, in order to provide a scale that allows a clear comparison between the BOMD and PIMD simulations. See Figure \ref{fig:gap_pimd_df2} for a more complete set of experimental results. } 
     \label{fig:C2c_gap_pivsmd}
\end{figure}

Having established the influence of NQEs and DFs on the finite-temperature structure, we now turn to their influence on the bandgap. % , and correlations we can make to the structure. 
Figure \ref{fig:C2c_gap_pivsmd} shows a comparison of the pressure dependence of the electronic bandgap for $C2c$ at $200$ K, for simulations using PBE. 
The bandgap is seen to be dramatically affected by NQEs, not only in its magnitude, but also in its pressure dependence. While calculations on static crystals result in bandgaps in reasonable agreement with experiment, the proper inclusion of NQEs leads to conditions where the bandgap closes at pressures as low as $p \jmmapprox 200$ GPa. Furthermore, for PIMD simulations using PBE, the bandgaps of all three structures close below $250$ GPa. Note that these results are in disagreement with experimental measurements \cite{Loubeyre02,Howie12,Zha12,Goncharov12}. 

\begin{figure}[ht]
     \includegraphics[scale=0.80]{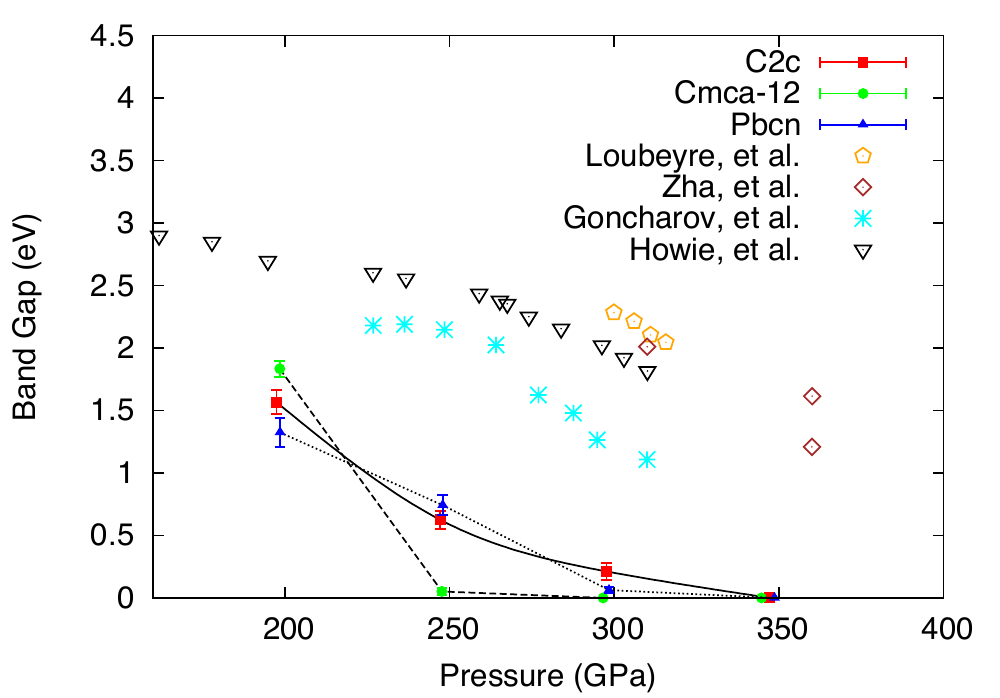}
     \caption{(Color online) Bandgap as a function of pressure. Filled symbols correspond to PIMD simulations with HSE, while empty symbols correspond to experimental results; lines represent guides to the eye. Red squares, green circles, and blue triangles are theoretical results for the $C2c$, $Cmca$-$12$ and $Pbcn$ structures, respectively. The experimental results correspond to: orange pentagrams, Loubeyre, \emph{et al.} \cite{Loubeyre02}; brown diamonds, Zha \emph{et al.}, \cite{Zha12}, cyan asterisks, Goncharov, \emph{et al.}, \cite{Goncharov12}, and downward black triangles, Howie, \emph{et al.}, \cite{Howie12}.   } 
     \label{fig:gap_pimd_hse}
\end{figure}

\begin{figure}[ht]
     \includegraphics[scale=0.80]{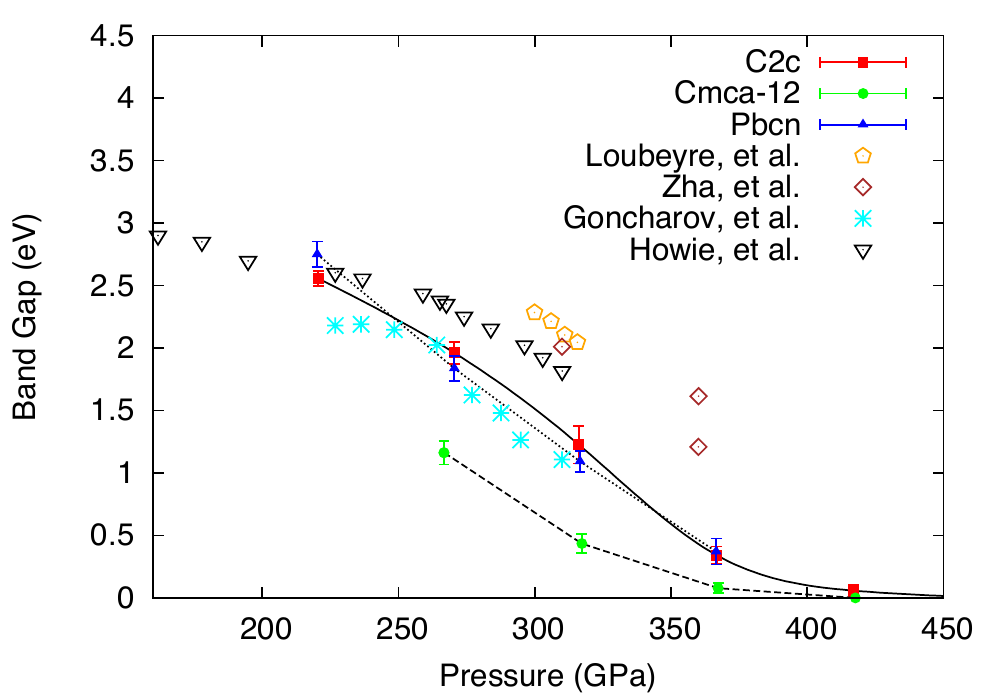}
     \caption{(Color online) Bandgap as a function of pressure. Filled symbols correspond to PIMD simulations with vdW-DF2. See the caption of Figure \ref{fig:gap_pimd_hse} for additional details.}
     \label{fig:gap_pimd_df2}
\end{figure}

While we have already established a strong influence of the DF, the same behavior is observed in all structures considered and regardless of the DF.  Figs.\ \ref{fig:gap_pimd_hse} and \ref{fig:gap_pimd_df2}  show a comparison of the pressure dependence of the bandgap for PIMD simulations performed with HSE and vdW-DF2; note that results using PBE are not shown, since, as just discussed, the bandgap goes to zero between 200 - 250 GPa in all structures. 
Some experimental results are also shown in these figures.
Several points are worth emphasizing. On the one hand, the use of HSE does not lead to a significant improvement compared to PBE, since, while the bandgaps
are finite, they are still considerably below the experimental ones. On the other hand, results obtained with vdW-DF2 agree very well, both the magnitude of the bandgap and its pressure dependence. 
These results suggest that C2c or Pbcn are more suitable structures in the investigated pressure range and at T=200K.

%--------------------------------------------------
\subsection{Energetics}
%--------------------------------------------------

While the previous discussion shows the dramatic influence of NQEs on the structural and optical
properties of the solid, it is important to emphasize that no attempt was made to correct our BOMD data for quantum effects,
as is typically done for other properties such as pressures and free energies. 
The common approach, especially in the study of high-pressure hydrogen,
is to use the quasi-harmonic approximation (QHA) for solids \cite{Born54} 
to incorporate ZPE corrections to an otherwise classical treatment of the nuclei. 
While this method is fairly accurate in many materials, leading to reasonable
predictions when compared to experiment \cite{Hickel12}, its applicability to 
light elements at high pressure is problematical. In the case of solid molecular hydrogen, 
the combination of orientational order, large amplitude fluctuations, and large anharmonic effects
(both classical and quantum) make the application of this approach questionable, when taking into account
the accuracy required for the correct determination of crystal structures \cite{Pickard07}. 

\begin{figure}[t]
     \includegraphics[scale=0.8]{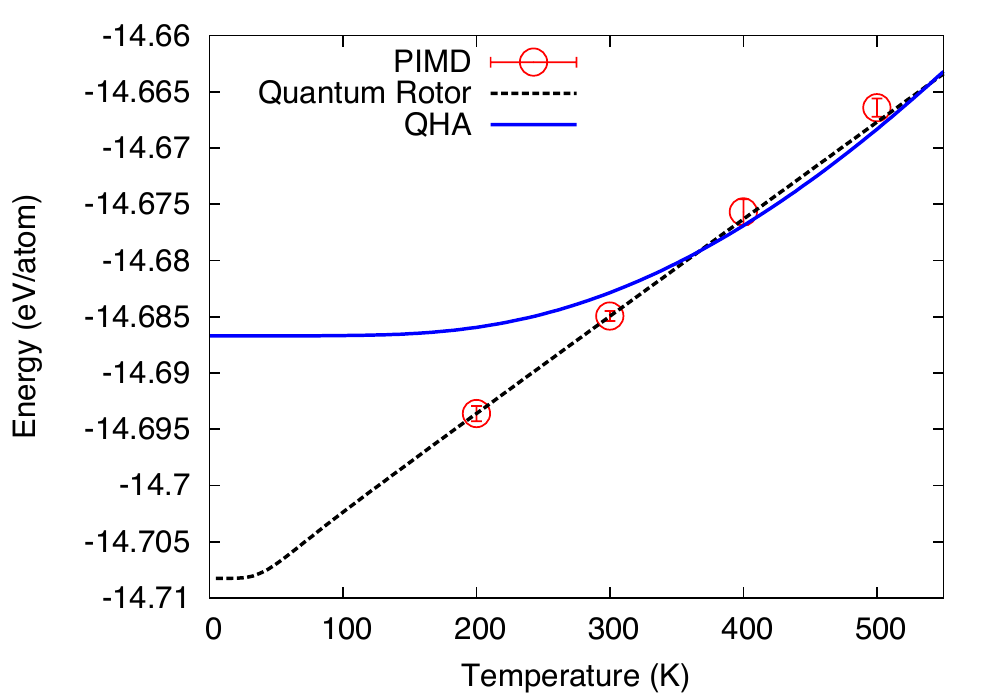} 
     \caption{(Color online) Internal energy of $Cmca$-$12$, as a function of temperature, at a density of $r_s \approx$ 1.38.
     Red squares correspond to PIMD simulations with vdW-DF2, the solid blue-curve corresponds to QHA results (also using vdW-DF2) and the dashed black-curve corresponds to the energy of an isolated quantum rotor, displaced to match the PIMD energy at $200$ K and with a rotational constant of an isolated hydrogen molecule of $\Theta_{Rot} \sim$ 87 K.} 
     \label{fig:qhavspi_cmca12}
\end{figure} 
Figure \ref{fig:qhavspi_cmca12} shows a comparison of the temperature dependence of the total energy of $Cmca$-$12$ at $\jmmapprox 265$ GPa between PIMD results using vdW-DF2 and the QHA.
A clear discrepancy in the overall temperature dependence is apparent. While the PIMD results show a clear
linear dependence in the regime studied, the QHA results show a very small effect up to $\jmmapprox 200$ K. The linear temperature-dependence of the energy in this regime can be understood by considering the fact that a hydrogen molecule has a rotational constant of approximately
$\Theta_{Rot} \approx 87$ K. This implies that the rotational component to the heat capacity acquires its 
classical value of $C_V^{Rot} = k_{B}$ above $\jmmapprox 100 $K in the case of distinguishable nuclei
considered here \footnote{Inclusion of Fermi statistics would modify the temperature range
where rotations behave quantum mechanically. This is the subject of future study.}.
Figure \ref{fig:qhavspi_cmca12} also shows the energy of an isolated quantum rotor (again assuming distinguishable
particles), displaced to match the PIMD energy at $200$ K. Notice the excellent agreement in the
resulting temperature dependence, supporting the fact that molecular rotations behave classically in this regime.
On the other hand, both intra-molecular and center-of-mass vibrations show strong quantum behavior in this regime.   
The results obtained with the QHA, on the other hand, do not capture this behavior, resulting in an incorrect 
temperature dependence of the energy. 
These results put into question the recent predictions of the transition
lines between Phases III and IV in molecular hydrogen based on QHA calculations\cite{Liu12}. 

\begin{figure}[t]
     \includegraphics[scale=0.7]{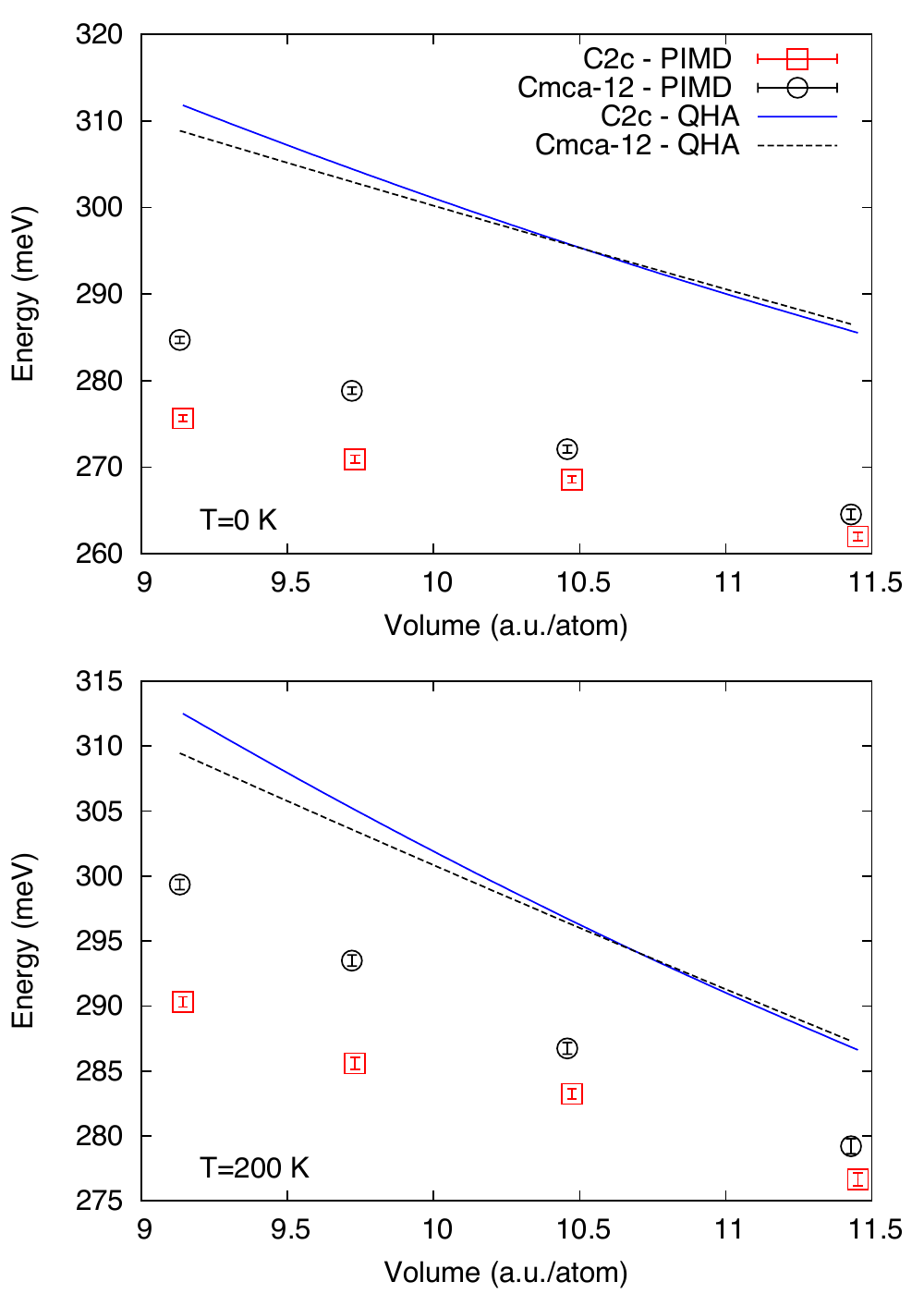}
     \caption{(Color online) Comparison between PIMD and the QHA of the thermal and quantum contribution to the internal energy of hydrogen in $C2c$ and $Cmca$-$12$ at $0$ K (top) and $200$ K (bottom).
     Symbols represent results with PIMD and lines the QHA. Note that in both cases, PBE was used.} 
     \label{fig:qhavspi_zpe}
\end{figure}

\begin{figure}[ht]
     \includegraphics[scale=0.75]{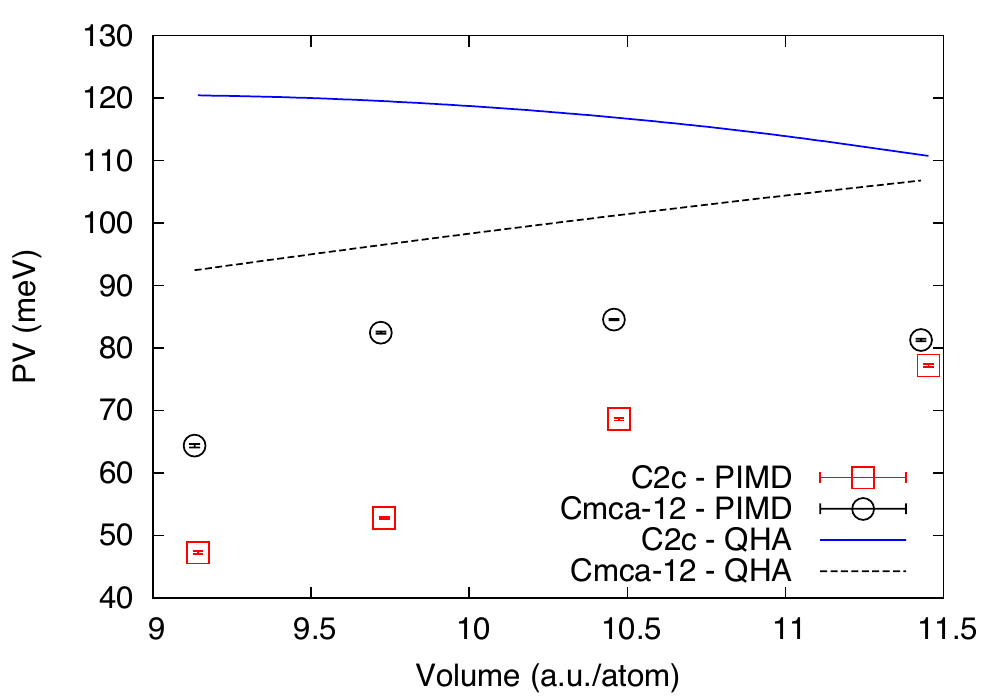}
     \caption{(Color online) Comparison between PIMD and the the QHA of the thermal and quantum contribution to the pressure component of the enthalpy in $C2c$ and $Cmca$-$12$. Symbols represent results with PIMD and lines with the QHA. Note that in both cases, PBE was used.} 
     \label{fig:qhavspi_zpp}
\end{figure}

Figures \ref{fig:qhavspi_zpe} and \ref{fig:qhavspi_zpp} show a comparison between PIMD
and the QHA of the combined thermal and quantum contributions to the internal energy 
and to the pressure component of the enthalpy ($PV$) using PBE.
Only results for $C2c$ and $Cmca$-$12$ are considered, 
since the ground state structure of $Pbcn$ displays imaginary phonons at $T=0$ K. 
For any thermodynamic quantity, $A$, we define its thermal and quantum contribution as 
$\Delta A_{X}(T) =  A_{X}(T) - A_\text{crystal}$, 
 $X$ represents either PIMD or QHA, and $A_\text{crystal}$ is the 
property calculated at $0$ K. 
Note also that the PIMD results at $0$ K are an estimation of the ZPE using the quantum rotor model,
as shown in Fig.\ \ref{fig:qhavspi_cmca12}.
While, as it has been demonstrated above, PBE is not expected to provide a good description of solid molecular hydrogen close to the metallization, these results provide a useful benchmark to measure the expected accuracy 
of the QHA.
Unfortunately though, there is a considerable discrepancy between
PIMD and the QHA, with the latter producing a large overestimate of the zero-point contribution
to both energies and enthalpies.
 
\begin{figure}[t]
     \includegraphics[scale=0.75]{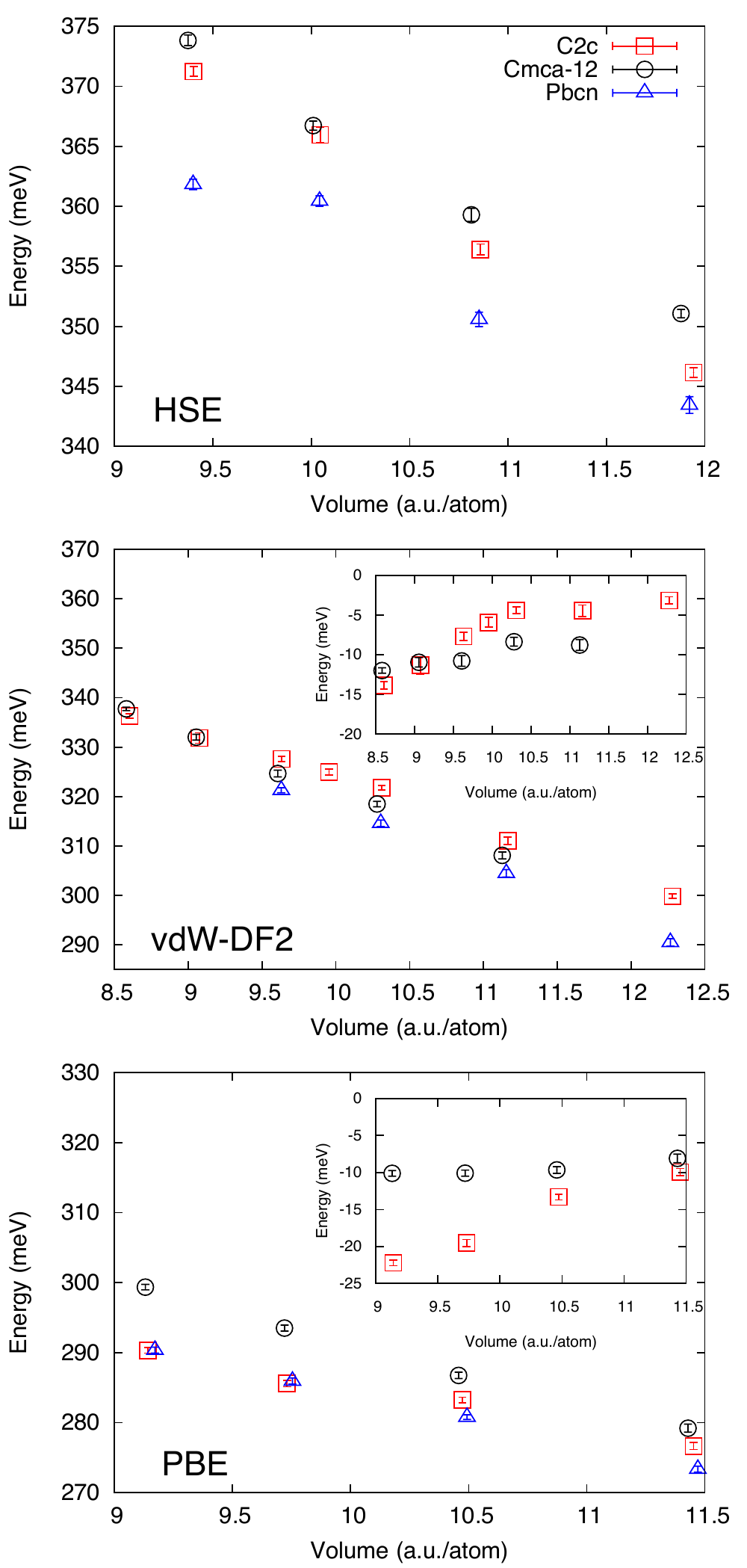}
     \caption{(Color online) Thermal and quantum contribution to the energy of hydrogen from PIMD simulations. From top to bottom the results correspond to: HSE, vdW-DF2, and PBE. Red squares, black circles, and blue triangles correspond to results for $C2c$, $Cmca$-$12$, and $Pbcn$, respectively. Insets show the error of the QHA, if it were to be employed. } 
     \label{fig:qhavspi_zpe_all}
\end{figure}

Figure \ref{fig:qhavspi_zpe_all} shows a comparison of the thermal and quantum contribution
to the internal energy of the structures from PIMD simulations, as a function of volume at $200$ K.
Note that, in addition, the error of the results if the QHA  were to be used is presented in the inset. Notice that the magnitude of the 
energy is dependent on DF, with PBE producing the smallest contribution and HSE the largest.
Also notice that the QHA error is dependent on both structure and functional, which eliminates
the possibility of combining results (e.g., ground-state energies and ZPE estimates) with different DFs to reduce the expense of the computations. In fact, the difference in energies between DFs is considerably larger than that between structures, for any given DF. 

\begin{figure}[t]
     \includegraphics[scale=0.82]{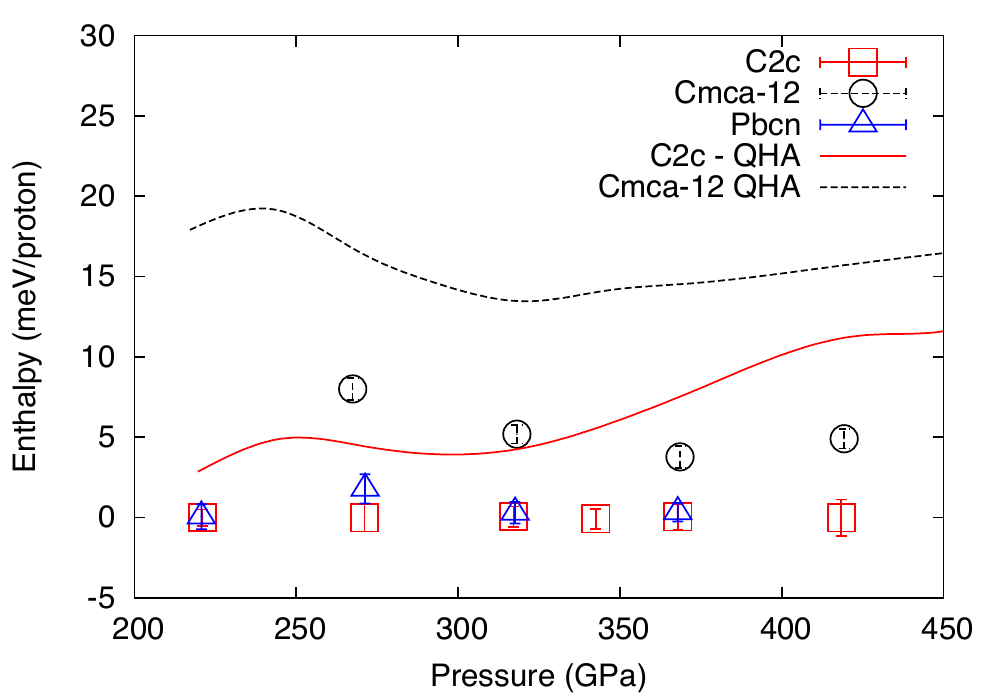}
     \caption{(Color online) Enthalpy of several structures of hydrogen at $200$ K, relative to a polynomial fit to that of $C2c$. Symbols represent PIMD simulations with vdW-DF2 and lines results from the QHA.} 
     \label{fig:pimd_enthalpy}
\end{figure}

While free-energy calculations have not been attempted in this work, the results presented nonetheless allow for a comparison of the enthalpy of solid at finite temperature, and also to assess the relative accuracy of the QHA with vdW-DF2.
Figure \ref{fig:pimd_enthalpy} shows the enthalpy of all three structures calculated using PIMD with vdW-DF2, along with results
using the QHA. Enthalpies are plotted relative to a polynomial fit to that of $C2c$. 
From these calculations, 
one cannot distinguish between the $C2c$ and the $Pbcn$ structure, since their enthalpies are within error bars of each other; also, one needs the relative entropies  to determine the stable structure. In addition, $Cmca$-$12$ 
has a higher enthalpy over the considered temperature and density range. 
The relative accuracy
of the QHA on the enthalpy is only around $5$--$10$ meV. While this is sufficient to establish a small list
of candidate structures, it is not enough to make a definite prediction of the phase diagram. 

According to figure \ref{fig:lind_vdW}, the $Cmca$-$12$ structure has lower Lindemann ratio and higher orientational order than the other two structures, a fact which signals a lower entropy for $Cmca$-$12$. Although a precise statement about the relative stability of the considered structures can only be based on a free-energy calculation, a lower entropy together with a higher enthalpy (see figure \ref{fig:pimd_enthalpy}) suggests the $Cmca$-$12$ structure to be less favorable than the others over the entire pressure range considered.

% therefore Can we claim on this basis that it will have the lowest entropy and therefore Cmca-12 will have higher free energy than the other structures??}  \MAM{I agree with Carlo's argument, is it worth adding a sentence? Energetically it is not more stable, so a lower lindeman ratio will not make it more stable, only less.}

%%%%%%%%%%%%%%%%%%%%%%%%%%%%%%%%%%%%%%%%%%%%%%%%%%
\section{Discussion}
\label{sec:discussion}
%%%%%%%%%%%%%%%%%%%%%%%%%%%%%%%%%%%%%%%%%%%%%%%%%%

The results above clearly show that in order to produce an accurate first-principles description of solid molecular hydrogen, especially close to metallization, not only does one need to take into account the quantum nature of the protons, but one also needs to go beyond standard semi-local descriptions of the electronic structure in DFT. The large ZPE of the protons increases the magnitude of 
molecular vibrations to the point that many structural features that are prominent in classical treatments end up being ``washed out''.
This can be understood by considering that the proper inclusion of NQEs
leads to nuclear kinetic energies of $\jmmapprox 1000$ K, even at temperatures of only $200$ K or lower,
which are not taken into account in classical simulations. In other words, 
the classical picture of a molecular bond at low temperatures is very different
from the quantum description; this can lead to artificially low displacements in such simulations.
While the magnitude of the ZPE can be estimated from the QHA,
its effects on the dynamical properties of the solid and its stability are not easy to 
estimate accurately with perturbative methods. 

The results also show the failure
of PBE in the correct description of molecular hydrogen close to dissociation. While it has been known 
for some time that PBE underestimates the bandgap in excited state calculations, there remains a wide-spread expectation
that the nuclear distribution produced by this DF should be nonetheless accurate. It is also believed that calculations with PBE should
produce an accurate description of the optical properties of hydrogen, as long as a more accurate
method, like hybrid functionals or GW, is used to calculate the gap. 
This has, in fact, been used to predict excellent agreement in calculated bandgaps \cite{Lenegue12,Goncharov12}.  
This is shown in figure 5 where a good agreement for the band gap of a perfect crystal predicted by PBE-DF AIRSS is obtained. However, this agreement is accidental as we have shown; even the use of HSE to sample the ionic distribution leads to 
poor results. Only by incorporating non-local correlation capable of
describing dispersion interactions provides a reasonable description of the optical properties of the solid obtained. Note that this is consistent with recent calculations performed in the liquid \cite{Morales13}. Unfortunately, 
the bandgaps of both $C2c$ and $Pbcn$ were very similar, which does not allow us to use these results to suggest any insight into the actual structure of Phase III or IV. This is a topic of future work.

Finally, the need to go beyond the QHA to estimate the impact of NQEs was demonstrated.
While the QHA significantly reduces the magnitude of the errors compared to a 
purely classical description of the protons, its accuracy is probably not enough to provide predictive 
results in some cases, e.g., the phase diagram. The relative energy differences between competing structures using the QHA is very small, with up to half a dozen structures separated by energies below 
10 meV  \cite{Pickard07}. While this approximation is certainly useful in identifying potential candidate structures, our results show that it is incapable
of producing results with an accuracy of $\jmmapprox 10$ meV, in some cases the error is larger.
 
Looking forward, since most of the work done to date in the field of high-pressure hydrogen has employed PBE and the results above suggest that vdW-DF2 produces the best description of solid molecular 
hydrogen among the DFs considered in this work, it is important 
not only to revisit the problem of structure prediction using vdW-DF2, 
but it is also important to rigorously assess its accuracy by comparison to more accurate many-body methods, such as quantum Monte Carlo \cite{Pierleoni06,Morales09,Morales10,Liberatore11}. In addition, a careful study that includes accurate free-energy methods for quantum
nuclei is needed in order to make accurate predictions of the correct structure ordering at finite temperature.

%%%%%%%%%%%%%%%%%%%%%%%%%%%%%%%%%%%%%%%%%%%%%%%%%%
\section{Summary and Conclusions}
\label{sec:concl}
%%%%%%%%%%%%%%%%%%%%%%%%%%%%%%%%%%%%%%%%%%%%%%%%%%%

In summary, the results shown in this article clearly illustrate two important limitations of current practices in the theoretical study of hydrogen: on the one hand, standard semi-local DFs like PBE lead to a poor description of high-pressure hydrogen \cite{Morales13}; and, on the other hand, the neglect of NQEs leads to predictions with very limited validity. While it has been recognized in the past  that both approximations should influence results in opposite directions, leading to a partial cancellation of errors, it is clear that the degree of cancellation depends on the thermodynamic conditions. Hence reliance on this cancellation leads to predictions of limited validity. %, and worse, to incorrect understanding when those predictions happen to accidentally agree with experiment. 

% ******************************
% ACKNOWLEDGMENTS
% ******************************
\begin{acknowledgments}
M.\ A.\ M.\ was supported by the U.S. Department of Energy at the Lawrence Livermore National Laboratory under Contract DE-AC52-07NA27344 and by LDRD Grant No.\ 13-LW-004. J.\ M.\ M.\ and D.\ M.\ C.\ were supported by DOE DE-FC02-06ER25794 and DE-FG52-09NA29456. C.\ P.\ was supported by the Italian Institute of Technology (IIT) under the SEED project Grant 259 SIMBEDD.
Computer time was provided by the US DOE INCITE program, Lawrence Livermore National Laboratory through the 7th Institutional Unclassified Computing Grand Challenge program and PRACE project n 2011050781.
\end{acknowledgments}

\end{document}